\RequirePackage{lineno}
\documentclass[aps,prd,twocolumn,superscriptaddress,showpacs,preprintnumbers,amsmath,amssymb]{revtex4}
\usepackage{graphicx} % Include figure files
\usepackage{dcolumn}  % Align table columns on decimal point
\usepackage{color} % 
\RequirePackage{xspace}
\usepackage{relsize}

\graphicspath{{ps}}

\begin{document}

%%% ***** Section 0 :  Title / Authors / Abstract

\vskip -1.0cm
\title{ \quad\\[1.0cm]
Search for a massive invisible particle $X^0$\\in $B^{+}\to e^{+}X^{0}$
and $B^{+}\to \mu^{+}X^{0}$ decays}
%%%%%%%%%%%%%%%%%%%%%%%%%%%%%%%%%%%%%%%%%%%%%%%%%%
%%% Paper:    B+ -> e+/mu+ X0
%%% Journal:  Physical Review D
%%% Contacts: C.S. Park (pcs4327@hanmail.net)
%%%           Y.J. Kwon (yjkwon63@yonsei.ac.kr)
%%% Non-responding authors or those who said NO are commented out.
%%% ====================================================================
%%% Click the RELOAD button on your web browser to see the updated file.
%%% ====================================================================
%%% Use \input{author} to insert this material into your latex file.
%%%%% Force institutions to appear in alphabetical order when typeset.
\noaffiliation
\affiliation{Aligarh Muslim University, Aligarh 202002}
\affiliation{University of the Basque Country UPV/EHU, 48080 Bilbao}
\affiliation{Beihang University, Beijing 100191}
%%%\affiliation{University of Bonn, 53115 Bonn}
\affiliation{Budker Institute of Nuclear Physics SB RAS, Novosibirsk 630090}
\affiliation{Faculty of Mathematics and Physics, Charles University, 121 16 Prague}
%%%\affiliation{Chiba University, Chiba 263-8522}
%%%\affiliation{Chonnam National University, Kwangju 660-701}
\affiliation{University of Cincinnati, Cincinnati, Ohio 45221}
\affiliation{Deutsches Elektronen--Synchrotron, 22607 Hamburg}
%%%\affiliation{University of Florida, Gainesville, Florida 32611}
%%%\affiliation{Department of Physics, Fu Jen Catholic University, Taipei 24205}
\affiliation{Justus-Liebig-Universit\"at Gie\ss{}en, 35392 Gie\ss{}en}
%%%\affiliation{Gifu University, Gifu 501-1193}
%%%\affiliation{II. Physikalisches Institut, Georg-August-Universit\"at G\"ottingen, 37073 G\"ottingen}
\affiliation{SOKENDAI (The Graduate University for Advanced Studies), Hayama 240-0193}
%%%\affiliation{Gyeongsang National University, Chinju 660-701}
\affiliation{Hanyang University, Seoul 133-791}
\affiliation{University of Hawaii, Honolulu, Hawaii 96822}
\affiliation{High Energy Accelerator Research Organization (KEK), Tsukuba 305-0801}
%%%\affiliation{Hiroshima Institute of Technology, Hiroshima 731-5193}
\affiliation{IKERBASQUE, Basque Foundation for Science, 48013 Bilbao}
%%%\affiliation{University of Illinois at Urbana-Champaign, Urbana, Illinois 61801}
\affiliation{Indian Institute of Technology Bhubaneswar, Satya Nagar 751007}
\affiliation{Indian Institute of Technology Guwahati, Assam 781039}
\affiliation{Indian Institute of Technology Madras, Chennai 600036}
%%%\affiliation{Indiana University, Bloomington, Indiana 47408}
\affiliation{Institute of High Energy Physics, Chinese Academy of Sciences, Beijing 100049}
\affiliation{Institute of High Energy Physics, Vienna 1050}
\affiliation{Institute for High Energy Physics, Protvino 142281}
%%%\affiliation{Institute of Mathematical Sciences, Chennai 600113}
\affiliation{INFN - Sezione di Torino, 10125 Torino}
\affiliation{J. Stefan Institute, 1000 Ljubljana}
\affiliation{Kanagawa University, Yokohama 221-8686}
\affiliation{Institut f\"ur Experimentelle Kernphysik, Karlsruher Institut f\"ur Technologie, 76131 Karlsruhe}
%%%\affiliation{Kavli Institute for the Physics and Mathematics of the Universe (WPI), University of Tokyo, Kashiwa 277-8583}
%%%\affiliation{Kennesaw State University, Kennesaw GA 30144}
\affiliation{King Abdulaziz City for Science and Technology, Riyadh 11442}
%%%\affiliation{Department of Physics, Faculty of Science, King Abdulaziz University, Jeddah 21589}
\affiliation{Korea Institute of Science and Technology Information, Daejeon 305-806}
\affiliation{Korea University, Seoul 136-713}
%%%\affiliation{Kyoto University, Kyoto 606-8502}
\affiliation{Kyungpook National University, Daegu 702-701}
\affiliation{\'Ecole Polytechnique F\'ed\'erale de Lausanne (EPFL), Lausanne 1015}
%%%\affiliation{Faculty of Mathematics and Physics, University of Ljubljana, 1000 Ljubljana}
\affiliation{Ludwig Maximilians University, 80539 Munich}
\affiliation{Luther College, Decorah, Iowa 52101}
\affiliation{University of Maribor, 2000 Maribor}
\affiliation{Max-Planck-Institut f\"ur Physik, 80805 M\"unchen}
\affiliation{School of Physics, University of Melbourne, Victoria 3010}
%%%\affiliation{Middle East Technical University, 06531 Ankara}
\affiliation{Moscow Physical Engineering Institute, Moscow 115409}
\affiliation{Moscow Institute of Physics and Technology, Moscow Region 141700}
\affiliation{Graduate School of Science, Nagoya University, Nagoya 464-8602}
\affiliation{Kobayashi-Maskawa Institute, Nagoya University, Nagoya 464-8602}
%%%\affiliation{Nara University of Education, Nara 630-8528}
\affiliation{Nara Women's University, Nara 630-8506}
\affiliation{National Central University, Chung-li 32054}
\affiliation{National United University, Miao Li 36003}
\affiliation{Department of Physics, National Taiwan University, Taipei 10617}
\affiliation{H. Niewodniczanski Institute of Nuclear Physics, Krakow 31-342}
%%%\affiliation{Nippon Dental University, Niigata 951-8580}
\affiliation{Niigata University, Niigata 950-2181}
\affiliation{University of Nova Gorica, 5000 Nova Gorica}
\affiliation{Novosibirsk State University, Novosibirsk 630090}
\affiliation{Osaka City University, Osaka 558-8585}
%%%\affiliation{Osaka University, Osaka 565-0871}
\affiliation{Pacific Northwest National Laboratory, Richland, Washington 99352}
%%%\affiliation{Panjab University, Chandigarh 160014}
%%%\affiliation{Peking University, Beijing 100871}
%%%\affiliation{University of Pittsburgh, Pittsburgh, Pennsylvania 15260}
%%%\affiliation{Punjab Agricultural University, Ludhiana 141004}
%%%\affiliation{Research Center for Electron Photon Science, Tohoku University, Sendai 980-8578}
%%%\affiliation{Research Center for Nuclear Physics, Osaka University, Osaka 567-0047}
%%%\affiliation{RIKEN BNL Research Center, Upton, New York 11973}
%%%\affiliation{Saga University, Saga 840-8502}
\affiliation{University of Science and Technology of China, Hefei 230026}
\affiliation{Seoul National University, Seoul 151-742}
%%%\affiliation{Shinshu University, Nagano 390-8621}
%%%\affiliation{Showa Pharmaceutical University, Tokyo 194-8543}
\affiliation{Soongsil University, Seoul 156-743}
\affiliation{University of South Carolina, Columbia, South Carolina 29208}
\affiliation{Sungkyunkwan University, Suwon 440-746}
\affiliation{School of Physics, University of Sydney, NSW 2006}
\affiliation{Department of Physics, Faculty of Science, University of Tabuk, Tabuk 71451}
\affiliation{Tata Institute of Fundamental Research, Mumbai 400005}
\affiliation{Excellence Cluster Universe, Technische Universit\"at M\"unchen, 85748 Garching}
\affiliation{Department of Physics, Technische Universit\"at M\"unchen, 85748 Garching}
\affiliation{Toho University, Funabashi 274-8510}
%%%\affiliation{Tohoku Gakuin University, Tagajo 985-8537}
\affiliation{Department of Physics, Tohoku University, Sendai 980-8578}
\affiliation{Earthquake Research Institute, University of Tokyo, Tokyo 113-0032}
\affiliation{Department of Physics, University of Tokyo, Tokyo 113-0033}
\affiliation{Tokyo Institute of Technology, Tokyo 152-8550}
\affiliation{Tokyo Metropolitan University, Tokyo 192-0397}
%%%\affiliation{Tokyo University of Agriculture and Technology, Tokyo 184-8588}
\affiliation{University of Torino, 10124 Torino}
%%%\affiliation{Toyama National College of Maritime Technology, Toyama 933-0293}
\affiliation{Utkal University, Bhubaneswar 751004}
\affiliation{CNP, Virginia Polytechnic Institute and State University, Blacksburg, Virginia 24061}
\affiliation{Wayne State University, Detroit, Michigan 48202}
\affiliation{Yamagata University, Yamagata 990-8560}
\affiliation{Yonsei University, Seoul 120-749}

  \author{C.-S.~Park}\affiliation{Yonsei University, Seoul 120-749} % Yonsei
  \author{Y.-J.~Kwon}\affiliation{Yonsei University, Seoul 120-749} % Yonsei
% \author{A.~Abdesselam}\affiliation{Department of Physics, Faculty of Science, University of Tabuk, Tabuk 71451} % Tabuk
  \author{I.~Adachi}\affiliation{High Energy Accelerator Research Organization (KEK), Tsukuba 305-0801}\affiliation{SOKENDAI (The Graduate University for Advanced Studies), Hayama 240-0193} % KEK
% \author{K.~Adamczyk}\affiliation{H. Niewodniczanski Institute of Nuclear Physics, Krakow 31-342} % Krakow
  \author{H.~Aihara}\affiliation{Department of Physics, University of Tokyo, Tokyo 113-0033} % Tokyo
% \author{S.~Al~Said}\affiliation{Department of Physics, Faculty of Science, University of Tabuk, Tabuk 71451}\affiliation{Department of Physics, Faculty of Science, King Abdulaziz University, Jeddah 21589} % Tabuk
% \author{K.~Arinstein}\affiliation{Budker Institute of Nuclear Physics SB RAS, Novosibirsk 630090}\affiliation{Novosibirsk State University, Novosibirsk 630090} % BINP
% \author{Y.~Arita}\affiliation{Graduate School of Science, Nagoya University, Nagoya 464-8602} % Nagoya
  \author{D.~M.~Asner}\affiliation{Pacific Northwest National Laboratory, Richland, Washington 99352} % PNNL
% \author{T.~Aso}\affiliation{Toyama National College of Maritime Technology, Toyama 933-0293} % Toyama
% \author{H.~Atmacan}\affiliation{Middle East Technical University, 06531 Ankara} % METU
% \author{V.~Aulchenko}\affiliation{Budker Institute of Nuclear Physics SB RAS, Novosibirsk 630090}\affiliation{Novosibirsk State University, Novosibirsk 630090} % BINP
  \author{T.~Aushev}\affiliation{Moscow Institute of Physics and Technology, Moscow Region 141700} % Lebedev
% \author{R.~Ayad}\affiliation{Department of Physics, Faculty of Science, University of Tabuk, Tabuk 71451} % Tabuk
% \author{T.~Aziz}\affiliation{Tata Institute of Fundamental Research, Mumbai 400005} % Tata
  \author{V.~Babu}\affiliation{Tata Institute of Fundamental Research, Mumbai 400005} % Tata
  \author{I.~Badhrees}\affiliation{Department of Physics, Faculty of Science, University of Tabuk, Tabuk 71451}\affiliation{King Abdulaziz City for Science and Technology, Riyadh 11442} % Tabuk
% \author{S.~Bahinipati}\affiliation{Indian Institute of Technology Bhubaneswar, Satya Nagar 751007} % IITB
  \author{A.~M.~Bakich}\affiliation{School of Physics, University of Sydney, NSW 2006} % Sydney
% \author{A.~Bala}\affiliation{Panjab University, Chandigarh 160014} % Panjab
% \author{Y.~Ban}\affiliation{Peking University, Beijing 100871} % Peking
% \author{V.~Bansal}\affiliation{Pacific Northwest National Laboratory, Richland, Washington 99352} % PNNL
  \author{E.~Barberio}\affiliation{School of Physics, University of Melbourne, Victoria 3010} % Melbourne
% \author{M.~Barrett}\affiliation{University of Hawaii, Honolulu, Hawaii 96822} % Hawaii
% \author{W.~Bartel}\affiliation{Deutsches Elektronen--Synchrotron, 22607 Hamburg} % DESY
% \author{A.~Bay}\affiliation{\'Ecole Polytechnique F\'ed\'erale de Lausanne (EPFL), Lausanne 1015} % Lausanne
% \author{I.~Bedny}\affiliation{Budker Institute of Nuclear Physics SB RAS, Novosibirsk 630090}\affiliation{Novosibirsk State University, Novosibirsk 630090} % BINP
% ...
% Ljubljana
% \author{N.~Zwahlen}\affiliation{\'Ecole Polytechnique F\'ed\'erale de Lausanne (EPFL), Lausanne 1015} % Lausanne
% \author{O.~Zyukova}\affiliation{Budker Institute of Nuclear Physics SB RAS, Novosibirsk 630090}\affiliation{Novosibirsk State University, Novosibirsk 630090} % BINP
\collaboration{The Belle Collaboration}

%%%%%%%%%%%%%%%%%%%%%%%%%%%%%%%%%%%%%%%%%%%%%%%%%%

\noaffiliation

%%% Paper:
%%% Journal:  Physical Review D
%%% Contacts: 
%%%
%%% Non-responding authors or those who said NO are commented out.
%%% ====================================================================
%%% Click the RELOAD button on your web browser to see the updated file.
%%% ====================================================================
%%% Use \input{author} to insert this material into your latex file.
%%%%% Force institutions to appear in alphabetical order when typeset.

\begin{abstract}

We present a search for a non-Standard-Model invisible particle $X^0$ in the mass 
range $0.1\textrm{-}1.8 \,{\rm GeV}/{c^2}$ in $B^{+}\to e^{+} X^{0}$ 
and $B^{+}\to \mu^{+} X^{0}$ decays.
The results are obtained from a $711~{\rm fb}^{-1}$ data sample
that corresponds to $772 \times 10^{6} B\bar{B}$ pairs,
collected  at the $\Upsilon(4S)$ resonance with the Belle detector at 
the KEKB $e^+ e^-$ collider.
One $B$ meson is fully reconstructed in a hadronic mode to determine the momentum
of the lepton of the signal decay in the rest frame of the recoiling partner $B$
meson. We find no evidence of a signal and set upper limits on the order of $10^{-6}$.

\end{abstract}

\maketitle
\tighten
%{\renewcommand{\thefootnote}{\fnsymbol{footnote}}}
%\setcounter{footnote}{0}

%\linenumbers

%%%%%% 1
Since their theoretical proposal by Pauli~\cite{Pauli} and the discovery
by Cowan {\it et al.}~\cite{Cowan}, neutrinos have played a crucial
role in developing and shaping the standard model (SM) of elementary
particle physics. Recent observation of neutrino oscillation~\cite{NO}
requires that they have non-zero masses. But in the minimal SM, there is
no mechanism for them to acquire non-zero mass.

Many new physics models beyond the SM introduce heavy neutrinos to
explain neutrino masses through the so-called seesaw mechanism~\cite{Seesaw}.
Moreover, these heavy neutrinos can help explain dark matter in
the universe. It is of great interest to search for heavy neutrino-like
particles. Such a heavy neutrino is an invisible particle, which we denote
$X^0$, and can be studied in $B^{+}$ decays to $l^+ X^0$~\cite{Charge},
where $l$ denotes an electron or muon.

\begin{figure}[htb]
\begin{center}
\includegraphics[width=0.2\textwidth]{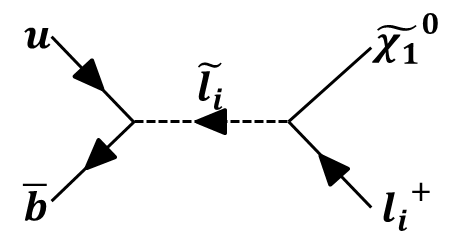}
\includegraphics[width=0.2\textwidth]{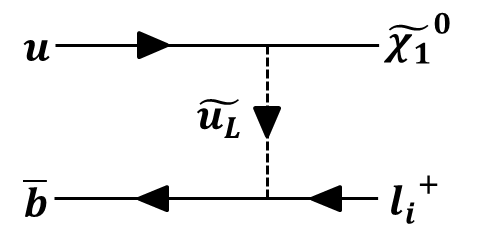}
\includegraphics[width=0.2\textwidth]{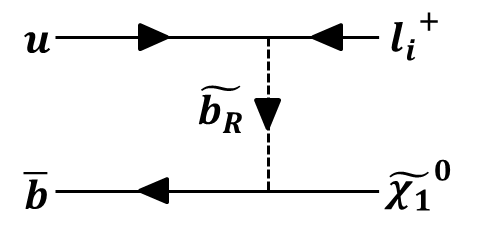}
\end{center}
\caption{Some Feynman diagrams to produce the lightest neutralino from $B$
meson decays in MSSM assuming $R$-parity violation.}
\label{fig:Feynman}
\end{figure}

There are further possibilities for the $X^0$ candidate in hypotheses of
new physics beyond the SM. One is sterile neutrinos in large extra
dimensions~\cite{nuLED} and in the neutrino minimal standard
model ($\nu$MSM) that incorporate the three light singlet right-handed
fermions~\cite{nuMSM}. Another option is the lightest supersymmetric
particle (LSP) in the minimal supersymmetric standard model (MSSM)~\cite{RSUSY}
assuming $R$-parity violation. If the $X^0$ is the LSP, it can be a neutralino
that is produced via the process shown in Fig.\ \ref{fig:Feynman}. If we
observe a particle $X^0$ that is significantly heavier than an SM neutrino,
it would indicate new physics.

%%%%%% 2

In this article, we report on searches for $B^{+} \to e^{+}X^{0}$ and
$B^{+} \to \mu^{+}X^{0}$ decays with an $X^0$ mass in the range
$0.1$ to $1.8\,{\rm GeV}/{c^2}$. The searches use
an $e^{+}e^{-}\to \Upsilon(4S)$ data sample of $711\,{\rm fb}^{-1}$
containing $772 \times 10^{6} \, B\bar{B}$ events produced by the
KEKB~\cite{KEKB} asymmetric $e^{+}e^{-}$ collider at $\sqrt{s} = 10.58 \,{\rm GeV}$,
which is at the $\Upsilon(4S)$ resonance, and recorded with the
Belle detector.

The Belle detector is a large-solid-angle magnetic spectrometer that
consists of a silicon vertex detector, a 50-layer central drift
chamber (CDC), an array of aerogel threshold \v{C}erenkov counters (ACC),  % <- \v{C}erenkov 2007.08
a barrel-like arrangement of time-of-flight scintillation counters, and
an electromagnetic calorimeter comprised of CsI(Tl) crystals (ECL)
located inside a super-conducting solenoid coil that provides a 1.5 T
magnetic field. An iron flux-return yoke located outside of the coil
is instrumented to detect $K_L^0$ mesons and to identify muons (KLM).
The detector is described in detail elsewhere~\cite{Belle}.

%We assume that the $X^0$ does not decay to observable particles inside the Belle detector.

We assume the $X^0$ is invisible and has a lifetime long enough to
escape from the Belle detector. Assuming a mean $X^0$ lifetime of
$10^{-6}$ seconds, fewer than 1\% of $X^0$ decay in the detector.
We search for a signal by exploiting the two-body decay kinematics of
$B^{+} \to l^{+}X^{0}$ decays. The magnitude $p_l^B$ of the momentum
of the charged lepton measured in the rest frame of the parent $B^+$
meson depends on the $X^0$ mass. The resolution of $p_l^B$ is affected
by the unknown direction of the parent $B^+$. To improve this resolution,
we fully reconstruct the other $B$ meson in the event in a hadronic
decay mode. For this reconstruction, an algorithm based on hierarchical
neural networks~\cite{EKP} is used. The charged $B$ meson, thus
reconstructed with 615 exclusive decay channels, is labeled $B_{\rm tag}$
and is used to constrain the kinematics of the signal $B$ meson. The
$B_{\rm tag}$ reconstruction quality for each candidate is denoted by a
variable $\it{o}_{\rm tag}$, which is the output from the neural network
algorithm. A $B_{\rm tag}$ candidate that is reconstructed with complete
certainty has $\it{o}_{\rm tag}={\rm 1}$ while one with no certainty has
$\it{o}_{\rm tag}={\rm 0}$.

When there are multiple $B_{\rm tag}$  candidates in an event, we
choose the candidate that has the largest $\it{o}_{\rm tag}$ value
from the hadronic tagging algorithm. We require
$\it{o}_{\rm tag}>{\rm 0.0025}$, for which the purity of the tagged
$B^+$ sample is 73\%; this falls to 56\% with a random selection of
the best $B_{\rm tag}$ candidate. To suppress combinatorially formed
$B_{\rm tag}$ candidates, we further require the following conditions
on the energy difference $\Delta E = E_{B_{\rm tag}} - \sqrt{s}/2$,
and the beam-energy-constrained mass
$M_{\rm bc}=\sqrt{ (s/4)/{c^4} - |\vec{p}_{B_{\rm tag}}|^{2}/{c^2}}$,
where $\vec{p}_{B_{\rm tag}}$ and $E_{B_{\rm tag}}$ are the
reconstructed momentum and energy, respectively, of the $B_{\rm tag}$
candidate in the center-of-mass (CM) frame:
$M_{\rm bc}>5.27~{\rm GeV}/c^2$ and $|\Delta E|<0.05~{\rm GeV}$.

The efficiency, $\epsilon_{\rm tag}$, of hadronic $B$ tagging is
initially determined by Monte Carlo (MC) simulation, then corrected
for a small data-MC difference by analyzing control sample modes composed
of the semileptonic $B^{+} \to \bar{D}^{(*)0}l^{+}\nu_{l}$ decays.
For $\bar{D}^{0}l^{+}\nu_{l}$, we consider only the $\bar{D}^0$
decays to $K^{+}\pi^{-}$, $K^{+}\pi^{-}\pi^{0}$, and $K^{+}\pi^{-}
\pi^{+}\pi^{-}$. For $\bar{D}^{*0}l^{+}\nu_{l}$, we use $\bar{D}^{*0}$
decays to $\bar{D}^{0}\pi^{0}$ and $\bar{D}^{0}\gamma$ with
$\bar{D}^{0}\to K^{+}\pi^{-}$.

We calculate the weighted average of the correction factors determined
from each control mode with their branching fractions as weights, as
described in Ref. \cite{TagCal}. After the correction, the efficiency
of the $B_{\rm tag}$ reconstruction is $0.17\%$ for $B^{+}\to e^{+}X^{0}$
and $0.18\%$ for $B^{+}\to \mu^{+}X^{0}$, with the relative uncertainty
of $\epsilon_{\rm tag}$ being 6.4\%~\cite{YOOK}.

%%%%%% 3

After removing particles used in the $B_{\rm tag}$ reconstruction, we
require that an event have only one charged track, that its charge be
opposite that of the $B_{\rm tag}$ and that its laboratory-frame momentum
exceed 1.0 ${\rm GeV}/c$. This charged track is required to satisfy
$|dz|<2.0\,{\rm cm}$ and $dr<0.5\,{\rm cm}$, where $|dz|$ and $dr$ are the
distances of closest approach to the interaction point along and
perpendicular to the beam axis.

We require that this charged track be identified as an electron or a muon.
Electrons are identified by means of a likelihood ratio based on the
following information: the ratio between the cluster energy in the ECL
and the track momentum from the CDC $(E/p)$, the specific ionization
$dE/dx$ in the CDC, the position and shower shape of the cluster in the
ECL and the response from the ACC. Muon identification uses the matching
information between the charged track and the KLM-hit positions as well as
the KLM penetration depth. With our track selection criteria, the electron
and muon efficiencies are over $90\%$ and their hadron misidentification
rates are below $0.5\%$ and $5\%$, respectively. A more detailed description
of the lepton identification can be found in Ref.~\cite{LID}.

The continuum background events ($e^+ e^- \to q\bar{q}$ with
$q=u,d,s,{\rm or}\,c$) are suppressed using the event shape difference
between $B\bar{B}$ and continuum events. In the CM frame, due to the
low momentum of the $B$ mesons, the event shape of a $B\bar{B}$ event
tends to be more spherical while the continuum backgrounds tend to
be more jet-like. To exploit this difference, we use the cosine of the
thrust angle, $\cos\theta_{\rm T}$, to suppress the continuum
backgrounds. Here, $\theta_{\rm T}$ is the angle between the thrust
axis of the $B_{\rm tag}$ and the momentum of the signal-side lepton
in the CM frame; the thrust axis is the direction that maximizes the
sum of the longitudinal momenta of the particles. We apply
$|\cos\theta_{\rm T}| < 0.9$ and $|\cos\theta_{\rm T}| < 0.8$ for
electron and muon candidates, relatively. The more stringent condition
is used for the muon due to its larger misidentification probability.

The remaining backgrounds, especially those with extra neutral particles
from the signal $B$ meson side, are suppressed by using the variable
$E_{\rm ECL}$, which is defined as the sum of the extra energy in the ECL
beyond that associated with the $B_{\rm tag}$ constituents and the
signal-side lepton. In calculating $E_{\rm ECL}$, we consider only clusters
with energies above 50 MeV in the barrel, 100 MeV in the forward endcap,
and 150 MeV in the backward endcap~\cite{Belle}. The higher thresholds in
the endcap regions reflect the more severe beam background in those
regions. We require $E_{\rm ECL}<0.5\, {\rm GeV}$ to enhance the signal.

\begin{figure}[htb]
\begin{center}
\includegraphics[width=0.4\textwidth]{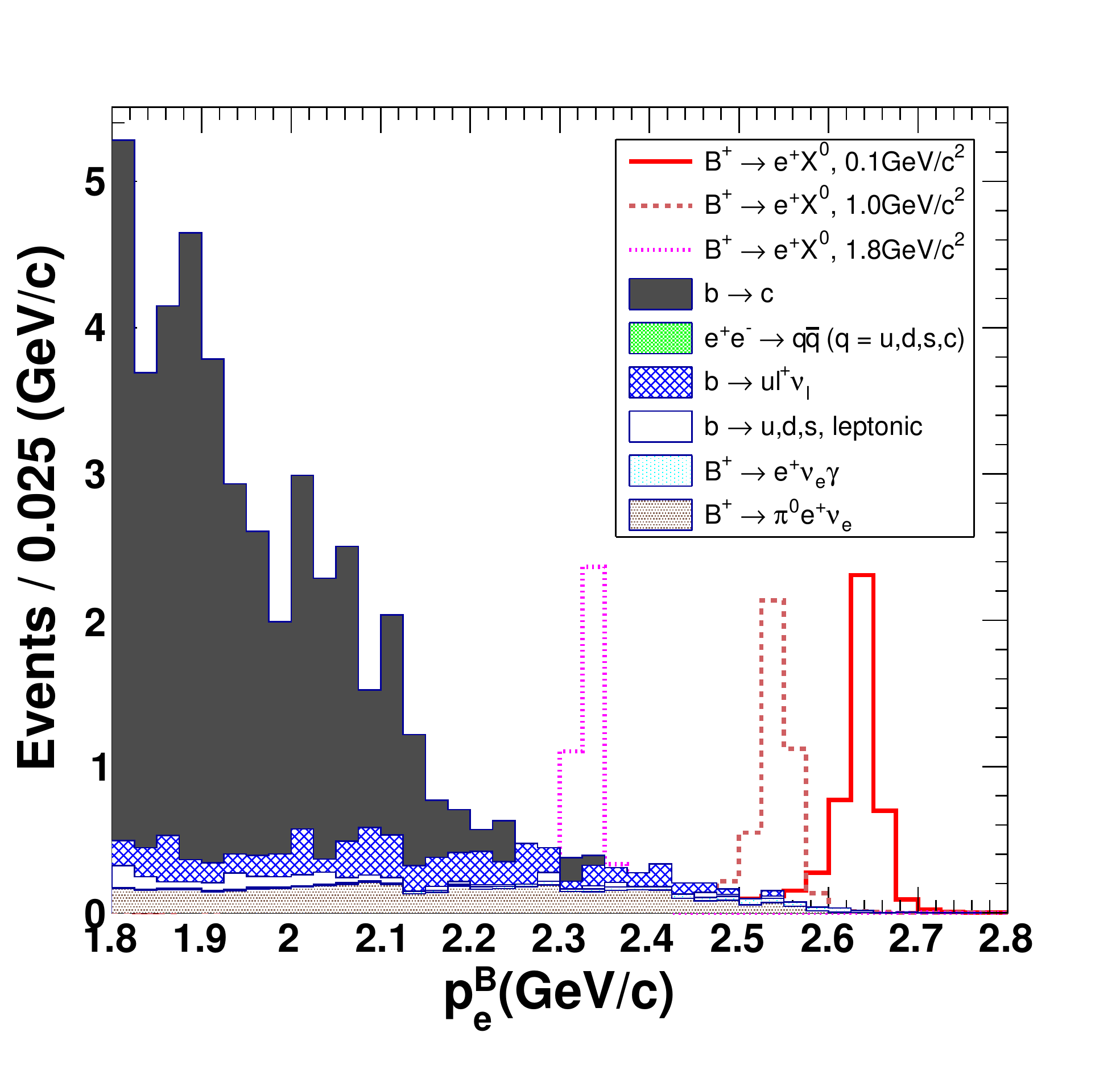}
\includegraphics[width=0.4\textwidth]{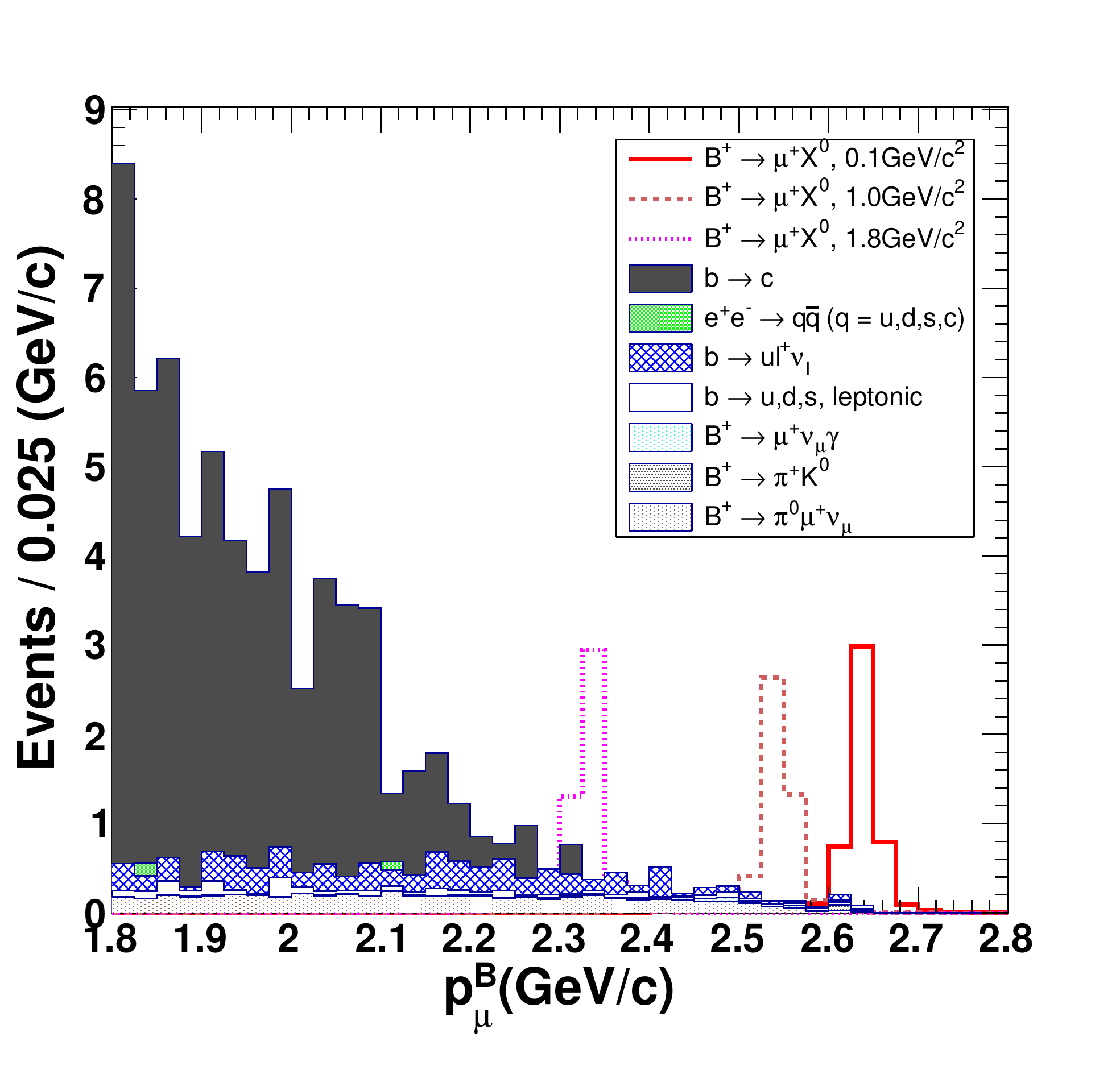}
\end{center}
\caption{$p_{l}^{B}$ MC distributions for $B^{+}\to e^{+}X^{0}$ (top) and
$B^{+} \to \mu^{+}X^{0}$ (bottom), where signal MC is arbitrary scaled.
The $e^+ e^- \to q\bar{q}$ background is negligible.
$B^{+} \to e^{+}\nu_{e}\gamma$, $B^{+} \to \mu^{+}\nu_{\mu}\gamma$ and
$B^{+} \to \pi^{+} K^{0}$ backgrounds become important for
$p_l^B > 2.5\,{\rm GeV}/c$.}
\label{fig:MC}
\end{figure}

We determine the signal yield using a fit to the $p_l^B$ distribution.
Figure \ref{fig:MC} shows the MC expectation for signal and background
for $p_l^B$ between $1.8 \, {\rm GeV}/c$ and $2.8 \, {\rm GeV}/c$.
The background level becomes increasingly significant as $p_l^B$ falls
below $2.3\, {\rm GeV}/c$.

As a result, we restrict our search to $M_{X^0}\leq 1.8\,{\rm GeV}/c^2$,
beyond which the search sensitivity is greatly degraded due to background.
For each assumed value of $M_{X^0}$, the $p_l^B$ signal region is
optimized based on the expected upper limit of the signal branching
fraction, which is estimated by MC simulation. Considering the width of
the so optimized signal regions of $p_l^B$ in Table \ref{table:Summary},
we perform the search in $0.1\,{\rm GeV}/c^2$ steps of $M_{X^0}$, whereby
the entire test region ($0.1\,{\rm GeV}/c^2 \leq M_{X^0} \leq 1.8\,{\rm GeV}/c^2$)
is covered without any gaps.

The number of expected background events in the $p_l^B$ signal region is
estimated by first performing a maximum likelihood fit to $p_l^B$ in the
region $1.8\,{\rm GeV}/c < p_l^B < 2.25\,{\rm GeV}/c$ (``sideband''),
where we expect very little contribution from the signal events for
$M_{X^0} < 1.8\,{\rm GeV}/c^2$. The fitted yield is then extrapolated to
the $p_l^B$ signal region, which is discussed in more detail below. To
fit the $p_l^B$ sideband, we consider the following sources of background:
continuum, $b\to c$ decays, semileptonic $b\to ul\nu$ decays, and other
rare and leptonic $B$-decay processes. The background distributions are
modelled by the probability density functions (PDFs), which are described
in Table \ref{table:Fit}. We do not consider continuum background in the
fitting because it is almost completely removed by our pre-selection.
Note that we utilize separate PDFs for the
$B^{+} \to l^{+}\nu_{l}\gamma$, $B^{+} \to \pi^{0}l^{+}\nu_{l}$, and
$B^{+} \to \pi^{+}K^{0}$ decays, as these modes show peaking behavior in
the $p_l^B$ distribution. The $B^{+} \to l^{+}\nu_{l}\gamma$ modes
(excluding taus), which have not been observed, could produce a
substantial yield of high-momentum leptons near the signal regions, so
we simulate them with dedicated large-sample-size MC. We use a branching
fraction of $2\times10^{-6}$ for $B^{+} \to e^{+}\nu_{e}\gamma$ and
$B^{+} \to \mu^{+}\nu_{\mu}\gamma$, which is lower than the recently
measured upper limit~\cite{Lnugam}. For $B^{+} \to \pi^{0} l^{+} \nu_l$,
$B^{+} \to \pi^{+}K^{0}$ and $B^{+} \to l^{+}\nu_l \gamma$,
high-statistics MC samples are produced with 300, 500, and 2500 times,
respectively, more integrated luminosity than the data. In the fit, only
the overall normalization is free and the relative yields of all background
modes are fixed based on the measured or assumed branching fractions. Finally,
the number of background events extrapolated in each signal region is
corrected by the data-MC difference. The correction factor is calculated
as the ratio of the number of events in the corresponding $p_l^B$ signal
region in the $E_{\rm ECL}$ sideband ($1.8\,{\rm GeV}/c<p_l^B < 3.0\,{\rm GeV}/c$,
$0.5\,{\rm GeV}<E_{\rm ECL}<2.0\,{\rm GeV}$) in data and in the MC sample.
The range of correction factors is 1.10 - 1.11 for the electron mode and
0.93 - 0.99 for the muon mode.

\begin{table*}[htb]
%\centering
%\begin{scriptsize}
\caption{Summary of upper limits at the 90\% CL.}
\label{table:Summary}
\begin{tabular}
{@{\hspace{0.5cm}}l@{\hspace{0.5cm}}
@{\hspace{0.5cm}}c@{\hspace{0.5cm}}
@{\hspace{0.5cm}}c@{\hspace{0.5cm}}
@{\hspace{0.5cm}}c@{\hspace{0.5cm}}
@{\hspace{0.5cm}}c@{\hspace{0.5cm}}
@{\hspace{0.5cm}}c@{\hspace{0.5cm}}}
\hline \hline
{} & {$p_{l}^{B}$ selection (${\rm GeV}/c$)} & {$\epsilon_{\rm s}$[\%]} & {$N_{\rm obs}$} & {$N_{\rm exp}^{\rm bkg}$} & {${\cal B}^{90}$}\\
\hline
{$M_{X^0}$} & \multicolumn{5}{c}{$B^{+} \to e^{+} X^{0}$ for $M_{X^0}$}\\
\hline
{0.1 ${\rm GeV}/c^2$} & {2.52-2.70} & {0.11} & {0} & {$0.36\pm0.13$} & {$< 2.4 \times 10^{-6}$}\\
{0.2} & {2.52-2.70} & {0.11} & {0} & {$0.36\pm0.13$} & {$< 2.4 \times 10^{-6}$}\\
{0.3} & {2.55-2.68} & {0.11} & {0} & {$0.21\pm0.13$} & {$< 2.6 \times 10^{-6}$}\\
{0.4} & {2.55-2.68} & {0.11} & {0} & {$0.21\pm0.08$} & {$< 2.7 \times 10^{-6}$}\\
{0.5} & {2.52-2.70} & {0.11} & {0} & {$0.36\pm0.08$} & {$< 2.5 \times 10^{-6}$}\\
{0.6} & {2.52-2.70} & {0.11} & {0} & {$0.36\pm0.13$} & {$< 2.5 \times 10^{-6}$}\\
{0.7} & {2.52-2.70} & {0.11} & {0} & {$0.36\pm0.13$} & {$< 2.4 \times 10^{-6}$}\\
{0.8} & {2.51-2.62} & {0.11} & {0} & {$0.37\pm0.12$} & {$< 2.5 \times 10^{-6}$}\\
{0.9} & {2.51-2.62} & {0.10} & {0} & {$0.37\pm0.12$} & {$< 2.6 \times 10^{-6}$}\\
{1.0} & {2.51-2.62} & {0.096} & {0} & {$0.37\pm0.12$} & {$< 2.8 \times 10^{-6}$}\\
{1.1} & {2.47-2.57} & {0.099} & {0} & {$0.58\pm0.18$} & {$< 2.4 \times 10^{-6}$}\\
{1.2} & {2.45-2.53} & {0.096} & {0} & {$0.61\pm0.19$} & {$< 2.5 \times 10^{-6}$}\\
{1.3} & {2.43-2.51} & {0.098} & {0} & {$0.72\pm0.22$} & {$< 2.3 \times 10^{-6}$}\\
{1.4} & {2.41-2.51} & {0.10} & {0} & {$0.97\pm0.30$} & {$< 2.0 \times 10^{-6}$}\\
{1.5} & {2.39-2.46} & {0.093} & {1} & {$0.85\pm0.27$} & {$< 4.8 \times 10^{-6}$}\\
{1.6} & {2.37-2.43} & {0.092} & {1} & {$0.84\pm0.27$} & {$< 4.9 \times 10^{-6}$}\\
{1.7} & {2.34-2.39} & {0.088} & {1} & {$0.85\pm0.28$} & {$< 5.1 \times 10^{-6}$}\\
{1.8} & {2.31-2.36} & {0.087} & {2} & {$1.01\pm0.34$} & {$< 7.1 \times 10^{-6}$}\\
\hline
{$M_{X^0}$} & \multicolumn{5}{c}{$B^{+} \to \mu^{+} X^{0}$ for $M_{X^0}$}\\
\hline
{0.1} & {2.58-2.68} & {0.12} & {1} & {$0.37\pm0.14$} & {$< 4.3 \times 10^{-6}$}\\
{0.2} & {2.58-2.68} & {0.12} & {1} & {$0.37\pm0.14$} & {$< 4.2 \times 10^{-6}$}\\
{0.3} & {2.58-2.68} & {0.12} & {1} & {$0.37\pm0.14$} & {$< 4.3 \times 10^{-6}$}\\
{0.4} & {2.58-2.68} & {0.12} & {1} & {$0.37\pm0.14$} & {$< 4.3 \times 10^{-6}$}\\
{0.5} & {2.58-2.68} & {0.11} & {1} & {$0.37\pm0.14$} & {$< 4.4 \times 10^{-6}$}\\
{0.6} & {2.58-2.68} & {0.11} & {1} & {$0.37\pm0.14$} & {$< 4.6 \times 10^{-6}$}\\
{0.7} & {2.56-2.63} & {0.11} & {0} & {$0.39\pm0.13$} & {$< 2.4 \times 10^{-6}$}\\
{0.8} & {2.54-2.61} & {0.11} & {1} & {$0.41\pm0.15$} & {$< 4.4 \times 10^{-6}$}\\
{0.9} & {2.52-2.60} & {0.11} & {1} & {$0.52\pm0.18$} & {$< 4.3 \times 10^{-6}$}\\
{1.0} & {2.49-2.58} & {0.11} & {1} & {$0.74\pm0.25$} & {$< 4.1 \times 10^{-6}$}\\
{1.1} & {2.49-2.58} & {0.12} & {1} & {$0.74\pm0.25$} & {$< 3.9 \times 10^{-6}$}\\
{1.2} & {2.48-2.53} & {0.10} & {0} & {$0.54\pm0.17$} & {$< 2.4 \times 10^{-6}$}\\
{1.3} & {2.45-2.50} & {0.10} & {0} & {$0.67\pm0.21$} & {$< 2.3 \times 10^{-6}$}\\
{1.4} & {2.42-2.48} & {0.11} & {2} & {$0.90\pm0.28$} & {$< 5.8 \times 10^{-6}$}\\
{1.5} & {2.40-2.47} & {0.11} & {5} & {$1.12\pm0.35$} & {$< 10.6 \times 10^{-6}$}\\
{1.6} & {2.37-2.42} & {0.10} & {4} & {$0.95\pm0.30$} & {$< 9.6 \times 10^{-6}$}\\
{1.7} & {2.34-2.39} & {0.10} & {1} & {$1.09\pm0.34$} & {$< 4.0 \times 10^{-6}$}\\
{1.8} & {2.31-2.37} & {0.11} & {1} & {$1.49\pm0.46$} & {$< 3.3 \times 10^{-6}$}\\
\hline \hline
\end{tabular}
%\end{scriptsize}
\end{table*}

\begin{table*}[htb]
\centering
\caption{Fit functions for background modes.}
\label{table:Fit}
\begin{tabular}
{@{\hspace{0.5cm}}l@{\hspace{0.5cm}}
@{\hspace{0.5cm}}c@{\hspace{0.5cm}}
@{\hspace{0.5cm}}c@{\hspace{0.5cm}}}
\hline \hline
{Background} & {$B^{+} \to e^{+} X^{0}$} & {$B^{+} \to \mu^{+} X^{0}$}\\
\hline
{$b\to c$} & {Gaussian} & {Gaussian}\\
{$b\to u l \nu_{l}$} & {Asymmetric Gaussian} & {Gaussian}\\
{$b\to u,d,s, {\rm leptonic}$} & {Exponential} & {Exponential + ARGUS~\cite{ARGUS}}\\
{$B^{+} \to l \nu_l \gamma$} & {Asymmetric Gaussian} & {Asymmetric Gaussian}\\
{$B^{+} \to \pi^{0} l \nu_l$} & {Asymmetric Gaussian + Gaussian} & {Asymmetric Gaussian + Gaussian}\\
{$B^{+} \to \pi^{+} K^{0}$} & {} & {Gaussian + Gaussian}\\
\hline \hline
\end{tabular}
\end{table*}

The signal branching fractions are obtained by the following equation:

\begin{equation}
{\cal B}(B^{+}\to l^{+}X^{0}) = \frac{N_{\rm obs}-N_{\rm exp}^{\rm bkg}}
{2\cdot \epsilon_{\rm s} \cdot N_{B^{+}B^{-}}}{\rm ,}
\end{equation}
where $N_{\rm obs}$ and $N_{\rm exp}^{\rm bkg}$ are the numbers of
observed and expected background events in the signal
region, $\epsilon_{\rm s}$ is the signal efficiency, and $N_{B^{+}B^{-}}$
is the number of $B^{+}B^{-}$ events.

To evaluate $\epsilon_{\rm s}$, signal MC samples are generated using
EvtGen~\cite{evtgen}, including final-state radiation using
PHOTOS~\cite{PHOTOS}. These samples are processed with a detector
simulation based on GEANT3~\cite{GEANT}. The signal efficiencies are
summarized in Table \ref{table:Summary}.

\begin{figure}[htb]
\begin{center}
\includegraphics[width=0.48\textwidth]{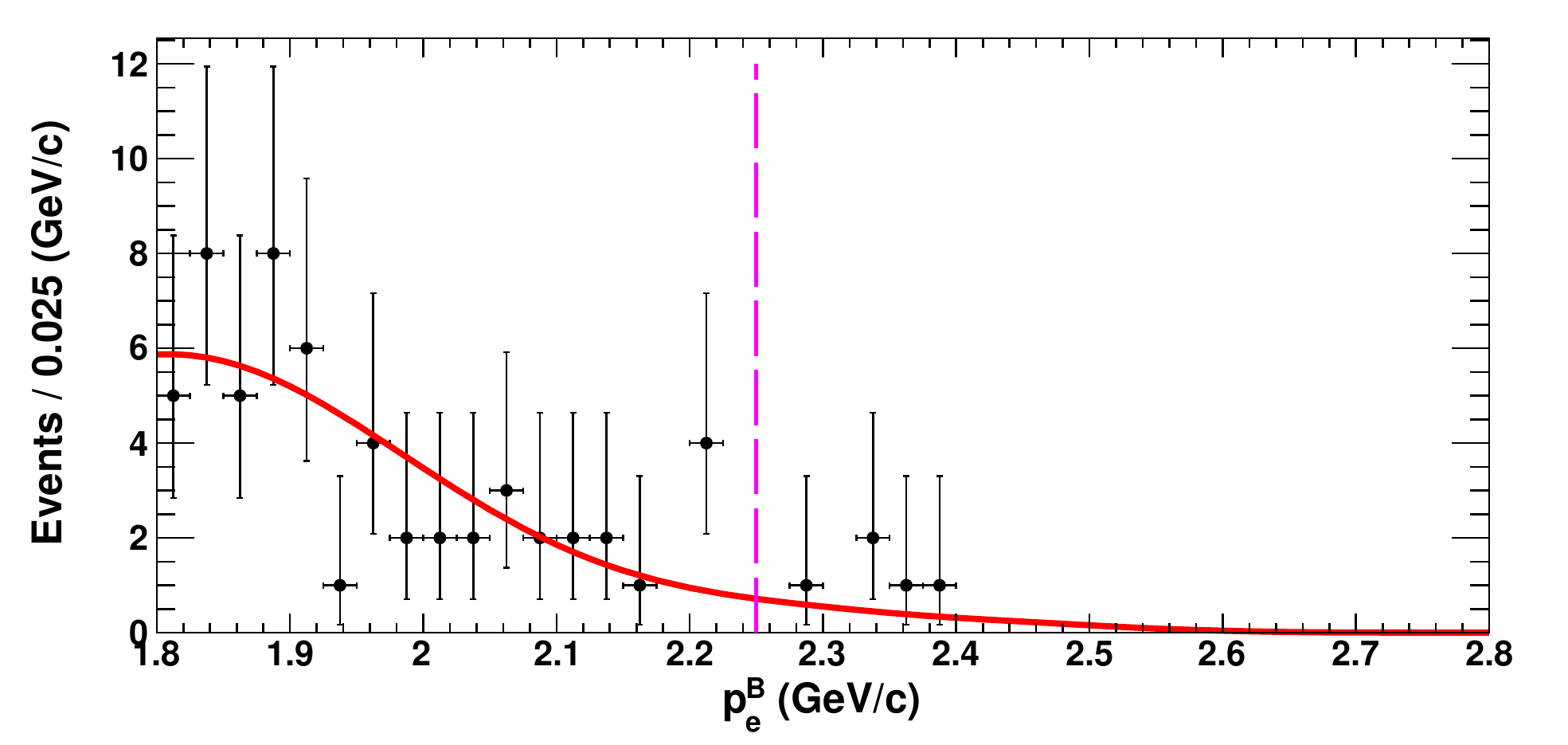}
\includegraphics[width=0.48\textwidth]{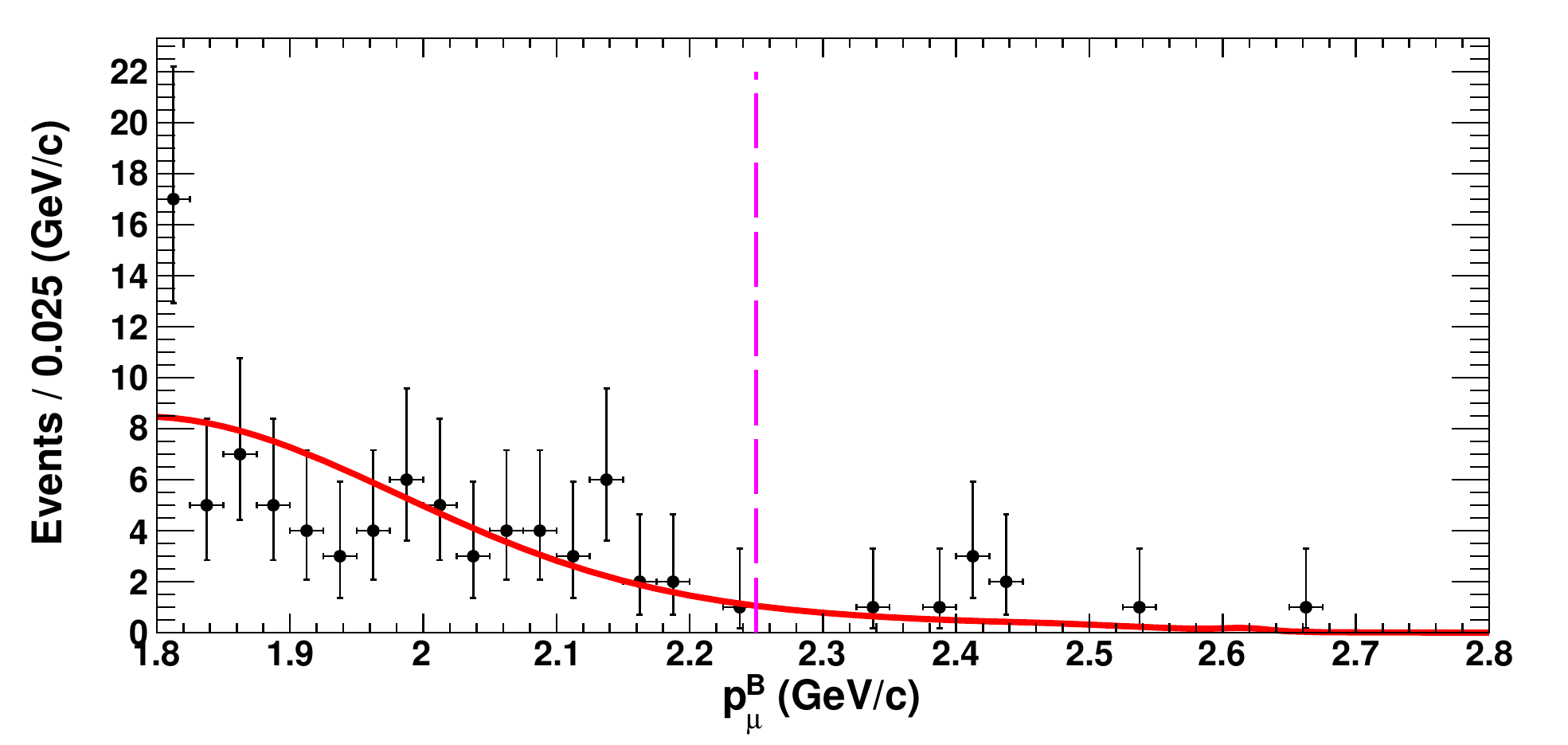}
\end{center}
\caption{$p_l^B$ data distributions for $B^{+} \to e^{+}X^{0}$ (top) and
$B^{+} \to \mu^{+}X^{0}$ (bottom), where the red curve indicates the
background expectation and the magenta dashed line indicates the upper
bound of the $p_l^B$ sideband.}
\label{fig:Data}
\end{figure}

Figure \ref{fig:Data} shows the $p_l^B$ distribution of the on-resonance
data. The fitted yield of background in the $p_l^B$ sideband of
on-resonance data is extrapolated to the signal region. The extrapolation
factor is determined from background MC samples.

The observed yields in the signal region are summarized in Table
\ref{table:Summary}. There is no signal excess for either mode in
any $M_{X^0}$ range. In the muon mode for $M_{X^0} = 1.5 \,{\rm GeV}/c^2$ ($1.6 \, {\rm GeV}/c^2$),
we find 5 (4) events in the $p_l^B$ signal region while we expect
$1.12\pm0.34$ ($0.95\pm0.29$) background events. The local $p$-value
of this yield, assuming a background-only hypothesis, is 0.60\%(1.59\%).
We obtain the 90\% confidence level (CL) upper limit of the signal yield
in each case by using the frequentist approach~\cite{FC} implemented in
the POLE (Poissonian limit estimator) program~\cite{POLE}, where the
systematic uncertainties are taken into account.

The systematic uncertainty consists of the multiplicative uncertainty
on $\epsilon_{\rm s}\cdot N_{B^+B^-}$ and the additive uncertainty on
the background. The multiplicative uncertainty is calculated from the
uncertainties on the number of $B^+B^-$ events, track finding and
lepton identification for the signal lepton, the $\epsilon_{\rm tag}$
correction, the $p_l^B$ shape, and the signal MC sample size.

A 1.8\% uncertainty is assigned for the uncertainty on the number of
$B$ mesons and the branching fraction of $\Upsilon(4S)\to B^+B^-$
\cite{PDG}. The track-finding uncertainty is estimated by comparing
the track-finding efficiency in data and MC, determining it in both
cases from the number of pions in the partially and fully reconstructed
$D^* \to \pi D^0$, $D^0 \to \pi\pi K_S^0$, $K_S^0 \to \pi\pi$ decay
chain. For the $p_l^B$ shape uncertainty, we use the 3.6\% uncertainty
from the $B^+ \to \bar{D}^0 \pi^+$ control sample study in the
$B^+ \to l^+ \nu_l$ search~\cite{YOOK} due to its similar kinematics.
The lepton identification uncertainty is estimated by comparing the
efficiency difference between data and MC using $\gamma\gamma \to l^+l^-$.
The multiplicative systematic uncertainties are summarized in Table
\ref{table:Syst}.

\begin{table}[htb]
\centering
\caption{Summary of multiplicative systematic uncertainties on
$\epsilon_{\rm s}\cdot N_{B^+B^-}$. The lepton identification and MC
statistical uncertainties depend on $M_{X^0}$ and are given as ranges.}
\label{table:Syst}
\begin{tabular}
{@{\hspace{0.5cm}}l@{\hspace{0.5cm}}
@{\hspace{0.5cm}}c@{\hspace{0.5cm}}
@{\hspace{0.5cm}}c@{\hspace{0.5cm}}}
\hline \hline
{Source} & {$B^{+} \to e^{+} X^{0}$} & {$B^{+} \to \mu^{+} X^{0}$}\\
\hline
{$N_{B^{+}B^{-}}$} & {1.8\%} & {1.8\%}\\
{Tracking} & {0.35\%} & {0.35\%}\\
{$\epsilon_{tag}$ correction} & {6.4\%} & {6.4\%}\\
{$p_{l}^{B}$ shape} & {3.6\%} & {3.6\%}\\
{Lepton ID} & {(1.0--1.1)\%} & {(0.8--0.9)\%}\\
{MC sample size} & {(1.8--2.0)\%} & {(1.8--1.9)\%}\\
\hline
{Total} & {7.9\%} & {7.8\%}\\
\hline \hline
\end{tabular}
\end{table}

The systematic uncertainties on the background estimation are
determined by considering the following sources: uncertainties
in the background PDF parameters, the branching fraction of the
background modes and the statistical uncertainty from the $p_l^B$
sideband. Each source is varied one at a time by its uncertainty
$(\pm 1\sigma)$ and the resulting deviations from the nominal
background yield are added in quadrature. For the branching
fraction uncertainties of the background modes, we use the
world-average values in Ref.~\cite{PDG} for $B^+ \to \pi^0l^+\nu_l$
and $B^+ \to \pi^+K^0$. For $B^+ \to l^+\nu_l\gamma$, a variation
of $\pm50\%$ is applied. For other modes, where an estimate of the
background level is not clearly available, a conservative branching
fraction uncertainty of $^{+100}_{-50}\%$ is assumed.

More than 95\% of $b \to c$ decays result in observed
$D^{(*)}l^+\nu_l$ final states, so we use their branching fraction
uncertainties~\cite{PDG}. The values of $N_{\rm exp}^{\rm bkg}$
and their uncertainties for both $B^+ \to e^+ X^0$ and
$B^+ \to \mu^+ X^0$ are listed in Table \ref{table:Summary}.

\begin{figure*}[htb]
\begin{center}
\includegraphics[width=0.46\textwidth]{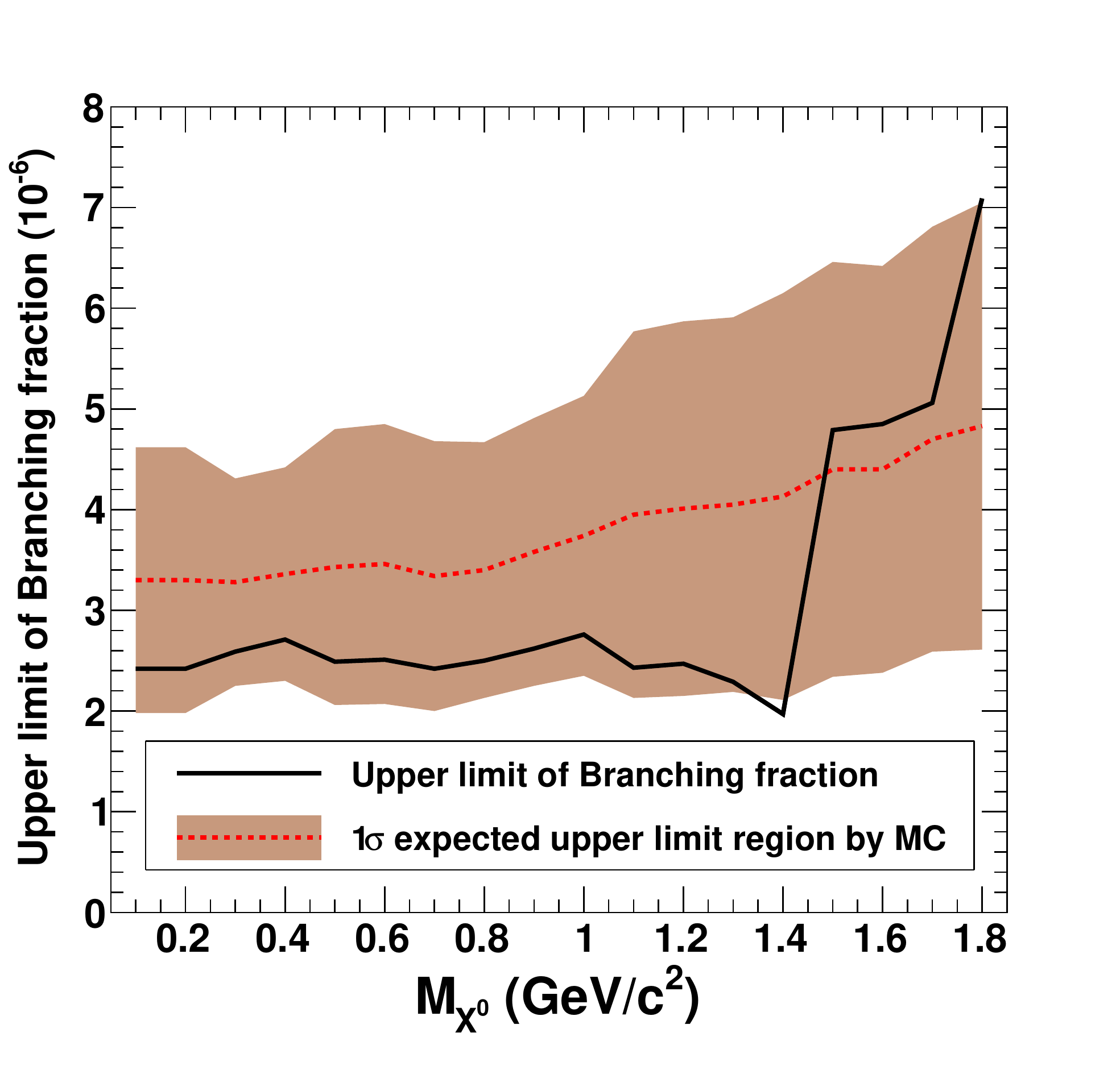}
\includegraphics[width=0.46\textwidth]{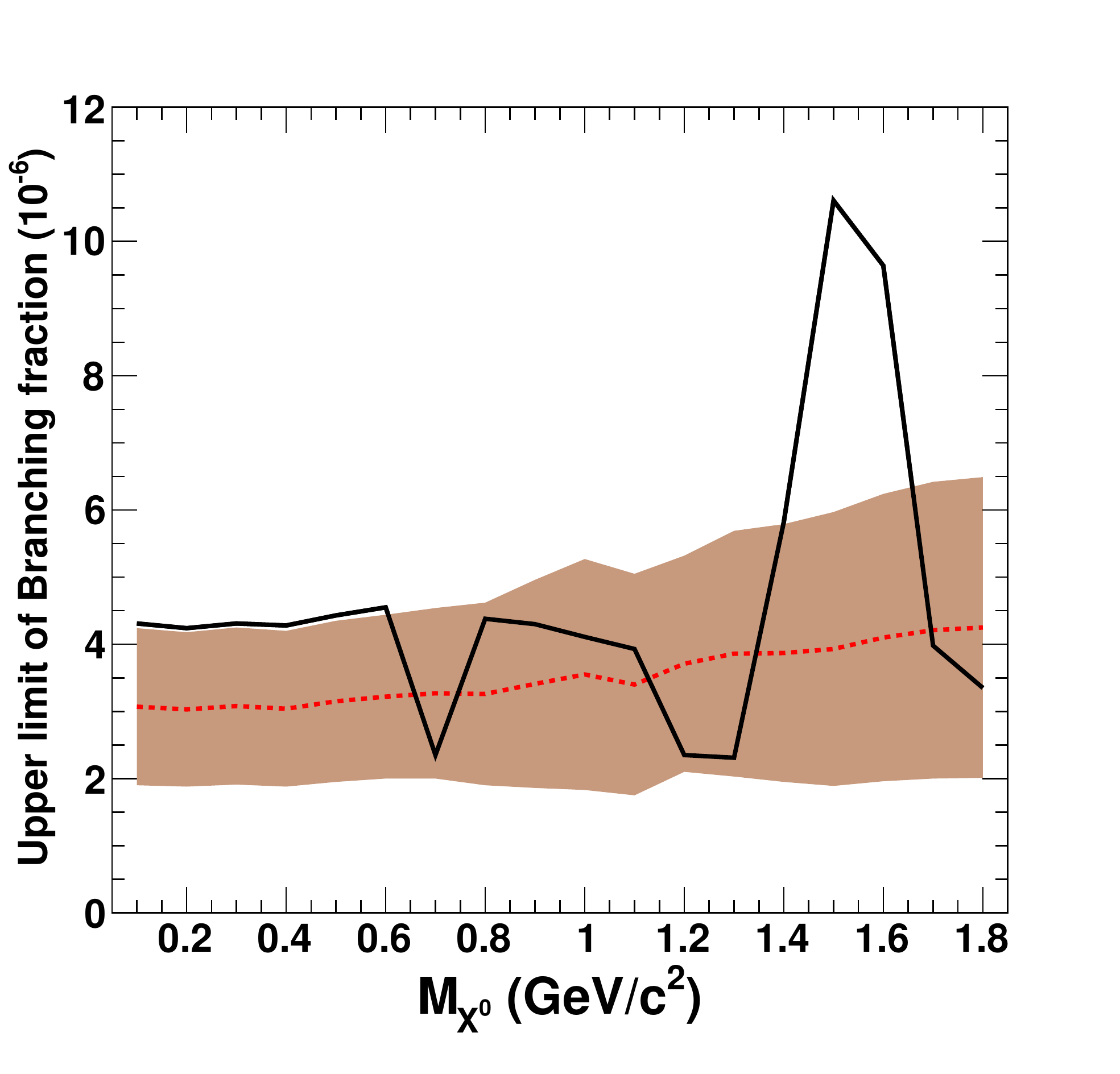}
\end{center}
\caption{The branching fraction upper limit as a function of $M_{X^0}$
and expected upper limit with 1$\sigma$ band; $e$ mode (left) and $\mu$
mode (right).}
\label{fig:BF}
\end{figure*}

Figure \ref{fig:BF} shows the expected number of background events
in the signal region as well as the obtained 90\% CL upper limits
of ${\cal B}(B^+ \to l^+ X^0)$ for each assumed value of $M_{X^0}$.
Table \ref{table:Summary} summarizes the $p_l^B$ signal region,
estimated background, signal efficiency, number of observed events,
and upper limit of the branching fraction at 90\% CL for each
assumed value of $M_{X^0}$ for both modes.

From the branching fraction upper limits, assuming $R$-parity
violation, we can set bounds on the MSSM-related parameter $\xi_l$
%\begin{widetext}
\begin{equation}
\begin{split}
\xi_l={\lambda^\prime}_{l13}^{2} \left( \frac{1}{2M_{\tilde{l}}^2}+\frac{1}{12M_{\tilde{u}_{L}}^2}+\frac{1}{6M_{\tilde{b}_{R}}^2} \right)^{2} \\
= \frac{8\pi{(m_{u}+m_{b})}^2 {\cal B}(B^{+}\to l^{+} X^{0})}{\tau_{B^+}{g^\prime}^{2} f_{B}^{2} m_{B^{+}}^{2} p_{l}^{B} (m_{B^+}^{2}-m_{l}^{2}-m_{X^0}^{2})}
\end{split}
\end{equation}
%\end{widetext}
where $\lambda'$ is a dimensionless $R$-parity-violating coupling
constant, $g'$ the weak coupling constant, $f_B$ the decay
constant of the $B^+$ meson, $m_{B^+}$ its mass, $p_l^B$ the
momentum of the $l^+$ in the $B$ rest frame, $m_{u}$ and $m_{b}$
the up and bottom quark mass, $m_{l}$ the charged lepton mass,
$m_{X^0}$ the neutralino mass, and $M_{\tilde{f}}$ the
sfermion mass that appears as an intermediate particle.
The range of upper bounds of $\xi_e$ is $4.1\times10^{-14}$ to
$1.7\times10^{-13}\,{\rm GeV}^{-4}c^{8}$ and on $\xi_\mu$ is
$4.2\times10^{-14}$ to $2.3\times10^{-13}\,{\rm GeV}^{-4}c^{8}$.

In summary, we obtain first upper limits for the branching fraction
of $B^+ \to e^+ X^0$ and $B^+ \to \mu^+ X^0$ for an $X^0$ mass range
$0.1\, {\rm GeV}/c^2$ to $1.8 \, {\rm GeV}/c^2$ using Belle's full
data set, where $X^0$ is assumed to leave no experimental signature.
For 18 assumed values of $M_{X^0}$ for both modes, upper limits of
branching fraction are found to be $O(10^{-6})$.

%***** Acknowledgments *****

%----------- Long version, for most papers ----------- 
We thank the KEKB group for the excellent operation of the
accelerator; the KEK cryogenics group for the efficient
operation of the solenoid; and the KEK computer group,
the National Institute of Informatics, and the 
PNNL/EMSL computing group for valuable computing
and SINET4 network support.  We acknowledge support from
the Ministry of Education, Culture, Sports, Science, and
Technology (MEXT) of Japan, the Japan Society for the 
Promotion of Science (JSPS), and the Tau-Lepton Physics 
Research Center of Nagoya University; 
the Australian Research Council;
Austrian Science Fund under Grant No.~P 22742-N16 and P 26794-N20;
the National Natural Science Foundation of China under Contracts 
No.~10575109, No.~10775142, No.~10875115, No.~11175187, No.~11475187
and No.~11575017;
the Chinese Academy of Science Center for Excellence in Particle Physics; 
the Ministry of Education, Youth and Sports of the Czech
Republic under Contract No.~LG14034;
the Carl Zeiss Foundation, the Deutsche Forschungsgemeinschaft, the
Excellence Cluster Universe, and the VolkswagenStiftung;
the Department of Science and Technology of India; 
the Istituto Nazionale di Fisica Nucleare of Italy; 
the WCU program of the Ministry of Education, National Research Foundation (NRF) 
of Korea Grants No.~2011-0029457,  No.~2012-0008143,  
No.~2012R1A1A2008330, No.~2013R1A1A3007772, No.~2014R1A2A2A01005286, 
No.~2014R1A2A2A01002734, No.~2015R1A2A2A01003280 , No. 2015H1A2A1033649;
the Basic Research Lab program under NRF Grant No.~KRF-2011-0020333,
Center for Korean J-PARC Users, No.~NRF-2013K1A3A7A06056592; 
the Brain Korea 21-Plus program and Radiation Science Research Institute;
the Polish Ministry of Science and Higher Education and 
the National Science Center;
the Ministry of Education and Science of the Russian Federation and
the Russian Foundation for Basic Research;
the Slovenian Research Agency;
Ikerbasque, Basque Foundation for Science and
the Euskal Herriko Unibertsitatea (UPV/EHU) under program UFI 11/55 (Spain);
the Swiss National Science Foundation; 
the Ministry of Education and the Ministry of Science and Technology of Taiwan;
and the U.S.\ Department of Energy and the National Science Foundation.
This work is supported by a Grant-in-Aid from MEXT for 
Science Research in a Priority Area (``New Development of 
Flavor Physics'') and from JSPS for Creative Scientific 
Research (``Evolution of Tau-lepton Physics'').


\begin{thebibliography}{99}

%\bibitem{Btolnu}
%N. ~Satoyama {\it et al.} (Belle Collaboration), Phys. Lett. B {\bf 647}, 67  (2007).
%\bibitem{Btotaunu}
%K.~Hara {\it et al.} (Belle Collaboration), Phys. Rev. D {\bf 82}, 072007(R) (2010); K.~Hara {\it et al.} (Belle Collaboration), Phys. Rev. Lett. {\bf 110}, 131801 (2013).
\bibitem{Pauli}
W. Pauli, in {\it Rapp. Septieme Conseil Phys. Solvay, Brussels 1933} (Gautier-Villars, Paris, 1934).
\bibitem{Cowan}
C. L. Cowan, Jr., F. Reines, F. B. Harrison, H. W. Kruse, and A. D. McGurie, Science {\bf 124}, 3212 (1956).
\bibitem{NO}
Y. Fukuda {\it et al.} (Super-Kamiokande Collaboration), Phys. Rev. Lett. {\bf 81}, 1158 (1998)
arXiv:hep-ex/9805021;
Q. R. Ahmad {\it et al.} (SNO Collaboration), Phys. Rev. Lett. {\bf 87}, 071301 (2001)
arXiv:nucl-ex/0106015.
\bibitem{Seesaw}
T. Yanagida, in Proc. of the Workshop on ``The Unified Theory and Baryon Number in the Universe", Tsukuba, Japan (1979) p.95;
M. Gell-Mann, P. Ramond and R. Slansky, in Supergravity, eds. P. van Nieuwenhuizen {\it et al.} (North-Holland, 1979), p. 315;
P. Minkowski, Phys. Lett. B {\bf 67} (1977) 421.
\bibitem{Charge}
Charge-conjugate decays are implied  throughout this paper unless otherwise stated.
\bibitem{nuLED}
K. Agashe, N. G. Deshpande, and G.-H. Wu, Phys. Lett. B {\bf 489}, 367 (2000)
arXiv:hep-ph/0006122.
\bibitem{nuMSM}
D. Gorbunov and M. Shaposhnikov, J. High Energy Phys. 10 (2007) 015
arXiv:0705.1729.
%\bibitem{nuMSM}
%A. Atre, T. Han, S. Pascoli, and B. Zhang, J. High Energy Phys. 05 (2009) 030.
\bibitem{RSUSY}
A. Dedes and H. Dreiner, Phys. Rev. D {\bf 65}, 015001 (2001)
arXiv:hep-ph/0106199
\bibitem{KEKB}
S.~Kurokawa and E.~Kikutani, Nucl. Instrum. Methods Phys. Res. Sect. A {\bf 499}, 1 (2003), and other papers included in this Volume;
T.Abe {\it et al.}, Prog. Theor. Exp. Phys. (2013) 03A001 and following articles up to 03A011.
\bibitem{Belle}
%A.~Abashian {\it et al.} (Belle Collaboration), Nucl. Instrum. Methods Phys. Res. Sect. A {\bf 479}, 117 (2002); also see detector section in J.~Brodzicka et al., Prog. Theor. Exp. Phys. (2012) 04D001.
A.~J.~Bevan {\it et al.}, Eur. Phys. J. C {\bf 74}, (2014) 3026
arXiv:1406.6311.
\bibitem{EKP}
M.~Feindt {\it et al.} (Belle Collaboration), Nucl. Instrum. Methods Phys. Res., Sect. A {\bf 654}, 432 (2011)
arXiv:1102.3876.
\bibitem{TagCal}
A.~Sibidanov {\it et al.} (Belle Collaboration), Phys. Rev. D {\bf 88}, 032005 (2013)
arXiv:1306.2781.
\bibitem{YOOK}
Y.~Yook {\it et al.} (Belle Collaboration), Phys. Rev. D {\bf 91}, 052016 (2015)
arXiv:1406.6356.
\bibitem{LID}
K.~Hanagaki {\it et al.}, Nucl. Instrum. Methods Phys. Res., Sect. A {\bf 485}, 490 (2002);
A.~Abashian {\it et al.}, Nucl. Instrum. Methods Phys. Res., Sect. A {\bf 491}, 69 (2002).
\bibitem{Lnugam}
A. Heller {\it et al.} (Belle Collaboration), Phys. Rev. D {\bf 91} (2015) 112009
arXiv:1504.05831.
\bibitem{ARGUS}
H.~Albrecht {\it et al.} (ARGUS Collaboration), Phys. Lett. B {\bf 241} (1990) 278.
\bibitem{evtgen}
D.~J.~Lange, Nucl. Instrum. Methods Phys. Res., Sect. A {\bf 462}, 152 (2001).
\bibitem{PHOTOS}
E.~Barberio and Z.~W\c{a}s, Comput. Phys. Commun. {\bf 79}, 291 (1994).
\bibitem{GEANT}
R.~Brun {\it et al.}, GEANT3.21, CERN Report DD/EE/84-1 (1984).
\bibitem{FC}
G.~J.~Feldman and R.~D.~Cousins, Phys. Rev. D {\bf 57}, 3873 (1998).
\bibitem{POLE}
J.~Conrad {\it et al.}, Phys. Rev. D {\bf 67}, 012002 (2003).
\bibitem{PDG}
%J.~Beringer {\it et al.} (Particle Data Group), Phys. Rev. D {\bf 86}, 010001 (2012) and 2013 partial update for the 2014 edition.
K.~A.~Olive {\it et al.} (Particle Data Group), Chin. Phys. C, {\bf 38}, 090001 (2014)

%\bibitem{SM}
%D.~Silverman and H.~Yao, Phys. Rev. D {\bf 38}, 214 (1988).
%\bibitem{CKM}
%N.~Cabibbo, Phys. Rev. Lett {\bf 10}, 531 (1963); M.~Kobayashi and T.~Maskawa, Prog. Theor. Phys. {\bf 49}, 652 (1973).
%\bibitem{PDG}
%J.~Beringer {\it et al.} (Particle Data Group), Phys. Rev. D {\bf 86}, 010001 (2012) and 2013 partial update for the 2014 edition.
%\bibitem{FBQCD}
%R.~J.~Dowdall {\it et al.} (HPQCD Collaboration), Phys. Rev. Lett. {\bf 110}, 222003 (2013).
%\bibitem{taube}
%K.~Hara {\it et al.} (Belle Collaboration), Phys. Rev. D {\bf 82}, 072007(R) (2010); K.~Hara {\it et al.} (Belle Collaboration), Phys. Rev. Lett. {\bf 110}, 131801 (2013).
%\bibitem{tauba}
%B.~Aubert {\it et al.} ($\textsc{Babar}$ Collaboration) Phys. Rev. D {\bf 76}, 052002 (2007); B.~Aubert {\it et al.} ($\textsc{Babar}$ Collaboration), Phys. Rev. D {\bf 77}, 011107(R) (2008).
%\bibitem{Belle2007}
%N.~Satoyama {\it et al.} (Belle Collaboration), Phys. Lett. B {\bf 647}, 67 (2007).
%\bibitem{Babar2009}
%B.~Aubert {\it et al.} ($\textsc{Babar}$ Collaboration), Phys. Rev. D {\bf 79}, 091101 (2009).
%\bibitem{2HDM}
%W.-S.~Hou, Phys. Rev. D {\bf 48}, 2342 (1993).
%\bibitem{MSSM}
%S.~Baek and Y.~G.~Kim, Phys. Rev. D {\bf 60}, 077701 (1999).
%\bibitem{LQ}
%H.~Georgi and S.~L.~Glashow, Phys. Rev. Lett. {\bf 32}, 438 (1974).
%\bibitem{MLFV}
%V.~Cirigliano, B.~Grinstein, G.~Isidori, and M.~B.~Wise, Nucl. Phys. B {\bf 728}, 121 (2005).
%\bibitem{FI}
%A.~Filipuzzi and G.~Isidori, Eur. Phys. J. C {\bf 64}, 55 (2009).
%\bibitem{tanbeta}
%{The parameter $\tan\beta$ is the ratio of the vacuum expectation values of the two Higgs fields; see A. Djouadi and J. Quevillon, JHEP \textbf{1310} (2013) 028.}
%\bibitem{Lepuniv}
%G.~Isidori and P.~Paradisi, Phys. Lett. B {\bf 639}, 499 (2006).
%\bibitem{Babar2008}
%B.~Aubert {\it et al.} ($\textsc{Babar}$ Collaboration), Phys. Rev. D {\bf 77}, 091104 (2008).
%\bibitem{Nuh}
%T.~Asaka, S.~Blanchet, and M.~Shaposhnikov, Phys. Lett. B {\bf 631}, 151 (2005); T.~Asaka and M.~Shaposhnikov, Phys. Lett. B {\bf 620}, 17 (2005).
%\bibitem{Belle}
%A.~Abashian {\it et al.} (Belle Collaboration), Nucl. Instrum. Methods Phys. Res. Sect. A {\bf 479}, 117 (2002); also see detector section in J.Brodzicka et al., Prog. Theor. Exp. Phys. (2012) 04D001.
%\bibitem{KEKB}
%S.~Kurokawa and E.~Kikutani, Nucl. Instrum. Methods Phys. Res. Sect. A {\bf 499}, 1 (2003), and other papers included in this Volume; T.Abe {\it et al.}, Prog. Theor. Exp. Phys. (2013) 03A001 and following articles up to 03A011.
%\bibitem{LID}
%K.~Hanagaki {\it et al.}, Nucl. Instrum. Methods Phys. Res., Sect. A {\bf 485}, 490 (2002); A.~Abashian {\it et al.}, Nucl. Instrum. Methods Phys. Res., Sect. A {\bf 491}, 69 (2002).
%\bibitem{GEANT}
%R.~Brun {\it et al.}, GEANT3.21, CERN Report DD/EE/84-1 (1984).
%\bibitem{EVTGEN}
%D.~J.~Lange, Nucl. Instrum. Methods Phys. Res., Sect. A {\bf 462}, 152 (2001).
%\bibitem{PHOTOS}
%E.~Barberio and Z.~W\c{a}s, Comput. Phys. Commun. {\bf 79}, 291 (1994).
%\bibitem{EKP}
%M.~Feindt {\it et al.}, Nucl. Instrum. Methods Phys. Res., Sect. A {\bf 654}, 432 (2011).
%\bibitem{Ulnu}
%A.~Sibidanov {\it et al.} (Belle Collaboration), Phys. Rev. D {\bf 88}, 032005 (2013).
%\bibitem{Barlow}
%R.~Barlow and C.~Beeston, Comp. Phys. Comm. {\bf 77} (1993) 219-228.
%\bibitem{Lnugam}
%G.~Korchemsky, D.~Pirjol and T.-M.~Yan, Phys. Rev. D {\bf 61}, 114510 (2000). 
%\bibitem{POLE}
%J.~Conrad {\it et al.}, Phys. Rev. D {\bf 67}, 012002 (2003).
%\bibitem{FC}
%G.~J.~Feldman and R.~D.~Cousins, Phys. Rev. D {\bf 57}, 3873 (1998).
%\bibitem{B2}
%T.~Abe {\it et al.} (Belle Collaboration), 	arXiv:1011.0352v1 [physics.ins-det] (2010).


\end{thebibliography}
\end{document}